\documentclass[10pt]{spie}  
\usepackage{tcolorbox}
\usepackage{tikz}
\usetikzlibrary{fit}% Package imports
\usepackage{amsmath,amsfonts,amssymb}
\usepackage{graphicx}
\usepackage[colorlinks=true, allcolors=blue]{hyperref}
\usepackage{subcaption} % For subfigures and captions
\usepackage{booktabs} % For better table formatting
\usepackage{xcolor} % For color enhancements in figures/tables
\title{Towards a Quantum-classical
Augmented Network}

\author[a]{Nitin Jha}
\affil[a]{Kennesaw State University, Marietta, GA, USA}
\author[a]{Abhishek Parakh}
\author[b]{Mahadevan Subramaniam}
\affil[b]{University of Nebraska Omaha, NE, USA}

\authorinfo{Further author information: Send correspondence to Nitin Jha at 
njha1@students.kennesaw.edu}

\begin{document} 
\maketitle

\begin{abstract}
In the past decade, several small-scale quantum key distribution networks have been established. However, the deployment of large-scale quantum networks depends on the development of quantum repeaters, quantum channels, quantum memories, and quantum network protocols. To improve the security of existing networks and adopt currently feasible quantum technologies, the next step is to augment classical networks with quantum devices, properties, and phenomena. To achieve this, we propose a change in the structure of the HTTP protocol such that it can carry both quantum and classical payload. This work lays the foundation for dividing one single network packet into classical and quantum payloads depending on the privacy needs. We implement logistic regression, CNN, LSTM, and BiLSTM models to classify the privacy label for outgoing communications. This enables reduced utilization of quantum resources allowing for a more efficient secure quantum network design. Experimental results using the proposed methods are presented. 
\end{abstract}


\section{INTRODUCTION}
\label{sec:intro} 
Recent decades have seen a great deal of interest in quantum information theory, which has led to advances in sensing, computing, and networking technologies \cite{Zhang2021DistributedQS, Aslam2023QuantumSF, Zhang2022FutureQC}. Under certain hardware configurations, quantum networks provide improved communication security \cite{renner2008}, and they are considered an essential component in future quantum internet development\cite{burr2022quantum}. Although there are certain obstacles \cite{NSA}, Quantum Key Distribution (QKD) is still a fundamental method for guaranteeing network security. Currently, research efforts are focused on improving QKD systems, creating near-ideal quantum hardware, and reducing the possibility of eavesdropping \cite{elliott2002building, yurke1984quantum, satoh2021attacking, parakh2018providing}. The DARPA network and other networks in Tokyo \cite{sasaki2011field} and throughout Europe \cite{peev2009secoqc} have shown that quantum networks are feasible in practice. However, these networks are small area networks, and developing a wide-spread quantum network becomes impractical due to the amount of noise present in current quantum devices. \cite{jha2024} 

\begin{figure}[h!]
    \centering
    \includegraphics[width=0.5\textwidth]{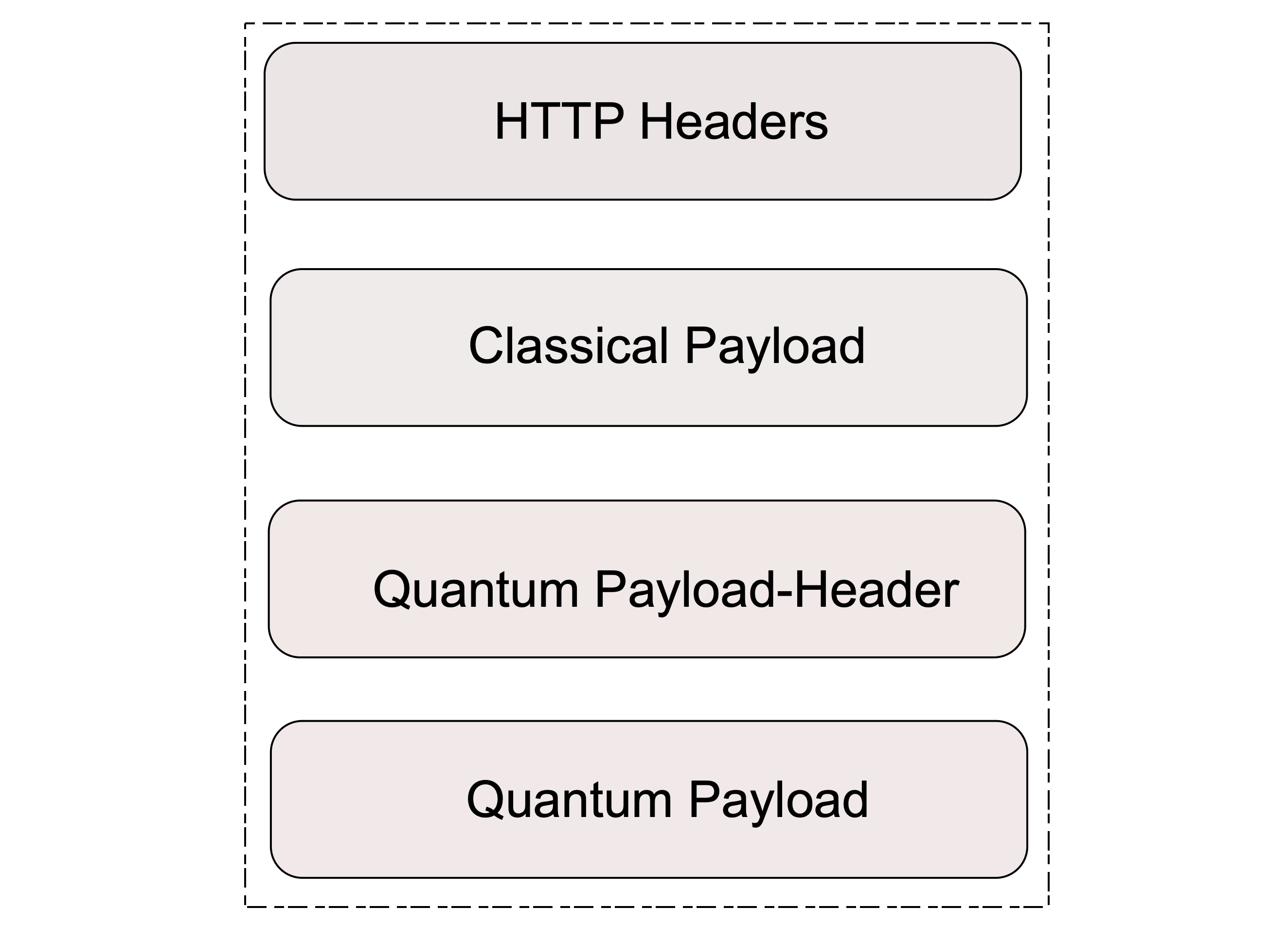}
    \caption{The proposed modified HTTP data-frame structure for a quantum-classical augmented network.}
    \label{fig:http-quant} 
\end{figure}

The challenges of implementing noise-free quantum channels and devices such as quantum repeaters are making large-scale quantum networks illusive \cite{repeaters}. To bridge the gap between large-scale classical networks and improve the security of the quantum networks, recently quantum augmented (or hybrid quantum-classical) networks have been proposed \cite{DARPA}. Using existing classical networking techniques and making them quantum-proof has been a long-standing idea in quantum communication. \cite{parakh2022quantum} Several of the quantum error correction methods exploit classical error correction techniques. \cite{jha2024joint} Several other researchers have looked at modification of existing protocols to support such hybrid networks\cite{10821224}.  For example, DiAdamo et al. \cite{PhysRev} define a hybrid data frame structure for transferring the quantum payload over a quantum-classical augmented network. In this design, the authors proposed to create a frame consisting of a quantum payload but with classical headers and trailers. Furthermore, the data frame discussed by the authors \cite{PhysRev} consists solely of a quantum payload. However, we propose an enhancement to create a hybrid data frame capable of carrying both classical and quantum payloads. This hybrid data frame can be directly integrated into modern networks, optimizing the use of quantum resources by reserving them for highly sensitive information. Specifically, we propose a modified HTTP data frame structure to accommodate both classical data and details about the quantum payload, as illustrated in fig(\ref{fig:http-quant}). The following are the functions of the layers mentioned in fig(\ref{fig:http-quant}),
\begin{enumerate}
    \item HTTP Headers: contains the general contents such as \textit{content type}, \textit{host name}, \textit{content length}, etc. 
    \item Classical Payload: contains the classical message.
    \item Quantum Payload-Header: contains information such as \textit{encoding scheme}, for eg., \textit{BB84} protocol, \textit{number of qubits used}, etc. 
    \item Quantum Payload: The actual quantum-encrypted data, basically the \textit{quantum state}. 
\end{enumerate}

\begin{figure*}[htbp]
    \centering
    \includegraphics[width=\linewidth]{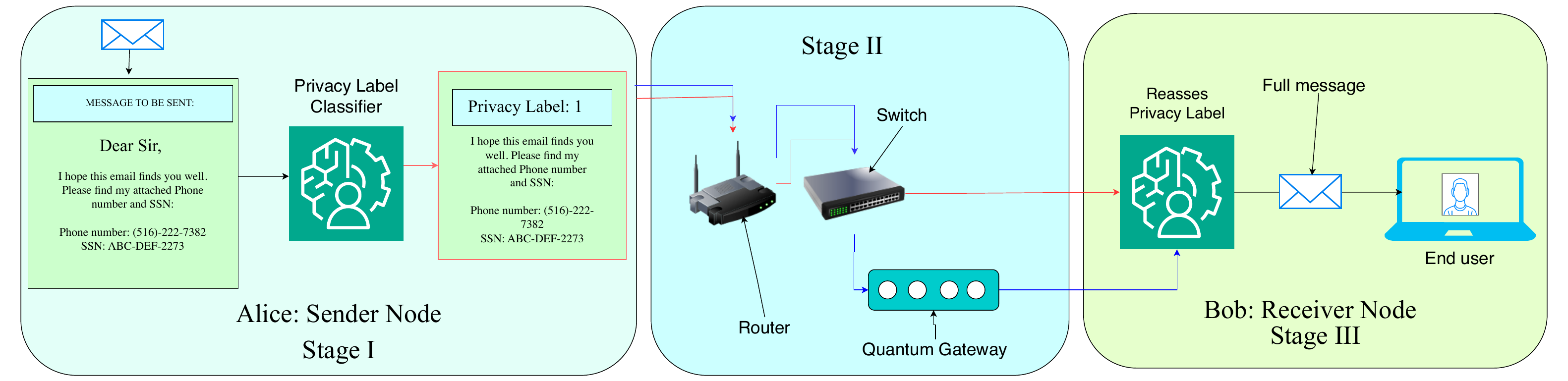}
    \caption{METHODOLOGY: Introduction of an ML model that parses a message and assigns a privacy label to each message, depending on the presence of any private information. Once any such quantum encryption passes through the quantum gateway and reaches the end user, this NLP model should be able to reassess the full message before the user can read it. Blue arrows indicates quantum payload, and red arrows indicates classical payload. We notice the presence of both arrows in Stage II, which is to signify that in a communication scenario, multiple packets would be transmitted. Red arrows shows the path of the packets with just classically encrypted information, and blue arrows shows the path taken by packets with quantum encrypted information.}
    \label{fig:nlp-struct}
\end{figure*}

Designing and implementing a full scale quantum network is quite resource expensive. The use of quantum resources and the vulnerability of quantum states to environmental noise makes this even more difficult. We, thus, introduce the idea of associating a privacy label to each outgoing packets at end-user nodes such that only the packets classified as private (denoted by privacy label of $1$, discussed more in detail in Sec[\ref{Sec:Dataset}]) are encoded in quantum states, thus saving up on expensive quantum resources. We aim to do this by utilizing several machine learning models which are discussed in more detail in later sections. This approach limits the use of quantum encryption by using classical encryption techniques for non-private communications. This approach ties in with the above described modified HTML protocol which is more suitable for basics of quantum augmented network. One important step in implementing the above modified HTML paradigm is to update the switch for a quantum-classical augmented network so that data packets can be directed to a quantum or a classical gateway, depending on the security needs of the message being transmitted.  For end-to-end encryption, we propose introducing a machine-learning algorithm that can parse the message and assign a privacy label, i.e., $0$ for messages without private contents, and $1$ for messages carrying private information. This is the main focus of this work. We tested several machine learning models (such as logistic regression, a basic CNN, LSTM, and BiLSTM) to assign privacy labels to each outgoing textual message (emails, texts, etc.). All the private messages can thus be encoded using quantum protocols. All the information regarding quantum encoding and the path to the receiver would go in the HTTP headers (both quantum and classical) for the switch to guide it through the quantum gateway. Fig(\ref{fig:nlp-struct}) describes the updated idea of a quantum-classical augmented network.

The updated quantum augmented network as described in fig(\ref{fig:nlp-struct}) can broadly be divided into three stages. Broadly, stage one is the sender node, stage two consists of any intermediate network connection, and stage three is the receiver nodes. A detailed description of the three stages are as follows, 
\begin{enumerate}
    \item \textbf{Stage I (Sender Node)}: The sender node (Alice) consists of the privacy label classifier which consists of several machine learning models which are discussed in detail in later sections. This is end-user application that will assess the privacy requirement of each outgoing communication, i.e., if the communication consists of any private information. This classifier assigns a privacy label of $0$ to any communication without any private information, and $1$ to any communication with private (or sensitive) information. If a communication is labelled to have private content, then it would be converted into a payload to be sent using a modified HTTP packet. This is essential for the next stage.

    \item \textbf{Stage II (Intermediate network nodes)}: The second stage is the network stage, i.e., when the packet leaves the origin and is trying to find it's destination. The packet structure is described in fig(\ref{fig:http-quant}). When the packet reaches a router, it's routed to the correct path using the addresses mentioned in the http-packet structure. When the packet reaches the switch, it'll assess if a packet contains any quantum payload or just has classical information. In the case of the quantum payload present, the packet is directed towards the quantum gateway. This is a theoretical construct where all the error corrections and decryption of the quantum states would happen.

    \item \textbf{Stage III (Receiver Node)}: Bob receives the measured classical information, i.e., the initial message sent by Alice. We include the same privacy classified model to reassess the privacy label of the received messages. The knowledge acquired at this end can be further used to backtrack and make the working of these classifiers better.

\end{enumerate}

This paper is structured as follows: Sec(\ref{Sec:related}) explores some of the related works that utilize machine learning techniques for privacy classification or threat detection. Most of these works are for classical networks. Sec(\ref{Sec:methods}) goes over the entire methodology used for this study. We look at the high-level architecture of the models used and a description of the dataset construction and other important statistics about it. Sec(\ref{Sec:result}) presents the results obtained in this study, and some discussions about it. Sec(\ref{Sec:conclusion}) concludes this work in which we use several machine learning models for the privacy classification of emails.  

\section{Related Works}
\label{Sec:related}
Numerous studies have previously employed the NLP paradigm to improve network security. Natural Language Processing (NLP) techniques such as Named Entity Recognition (NERs) were employed in one study by Kushwaha et al.,\cite{NLP-encrypt} to provide selective encryption for texts sent over mobile networks that was demonstrated to be more resource-efficient than existing methods. NLP models were employed in a different study by Yang et al. \cite{mal-NLP} to identify incoming malicious encrypted communications. As a result, we can create an NLP algorithm that can parse the message and identify the sections that require quantum-based encryption and the traditionally encrypted sections. For this format, quantum secure direct communication methods are perfect, like the three-stage QKD protocol \cite{kak2006}. Without previously establishing a key, this protocol can be used to deliver encrypted quantum states containing classical information directly. This is consistent with our proposed redesigned HTTP stack with both quantum and classical payload. The final user will receive a single information packet composed of classical and quantum components thanks to this NLP technique. Fig(\ref{fig:nlp-struct}) describes this complete process. In their study Shejwalkar et al.\cite{shejwalkar2021membership} discuss membership inference attacks (MIAs) against NLP classification models, highlighting the privacy risks related to text classification work. It compares sample-level MIAs, where an adversary tries to determine if a single data sample was a part of the training data, with user-level MIAs, which focus on the privacy of whole users. The paper demonstrates new and more effective user-level MIAs compared to existing sample-level attacks. By evaluating multiple NLP algorithms and datasets, the study draws attention to the grave privacy breaches that could arise from user-level MIAs.

\section{Network Architecture}
This work focuses on integrating the idea of having a privacy classifier directly into the communication network architecture. We already gave a high-level overview of how the privacy classifier serves the purpose of quantum augmented networks in fig(\ref{fig:nlp-struct}), now we go into the details of the architecture used at the sender and receiver's end. We start with a definition of single fiber idea, i.e., using one single fiber to transmit both classical and quantum information by using Wavelength Division Multiplexing (WDM).

\begin{itemize}
    \item \textbf{One Fiber}: Carries both classical data (TCP/IP) and quantum photons (QKD).
    \item \textbf{WDM Couplers}: Split/merge \emph{quantum wavelength} ($\lambda_q$) and \emph{classical wavelength} ($\lambda_c$).
    \item \textbf{Quantum Device}: Handles single-photon transmissions and key exchange. Required for private emails.
    \item \textbf{Router / NIC}: Handles classical IP traffic and passes Q-HTTP requests.
\end{itemize}

\begin{tcolorbox}[title=Example Q-HTTP Exchange, colback=white]
\small
\textbf{Request (from Alice):}
\begin{verbatim}
Q-SEND /secure-email HTTP/1.1
Host: bob.example.com
Q-Encryption-Mode: quantum
X-Quantum-Channel-ID: 1
Content-Type: application/octet-stream

[Encrypted Email Body (using QKD-derived key)] (Assuming quantum encryption entails that)
\end{verbatim}

\textbf{Response (from Bob):}
\begin{verbatim}
HTTP/1.1 200 OK
Q-Quantum-Status: success

[Optional: Status information, key index, etc.]
\end{verbatim}
\end{tcolorbox}

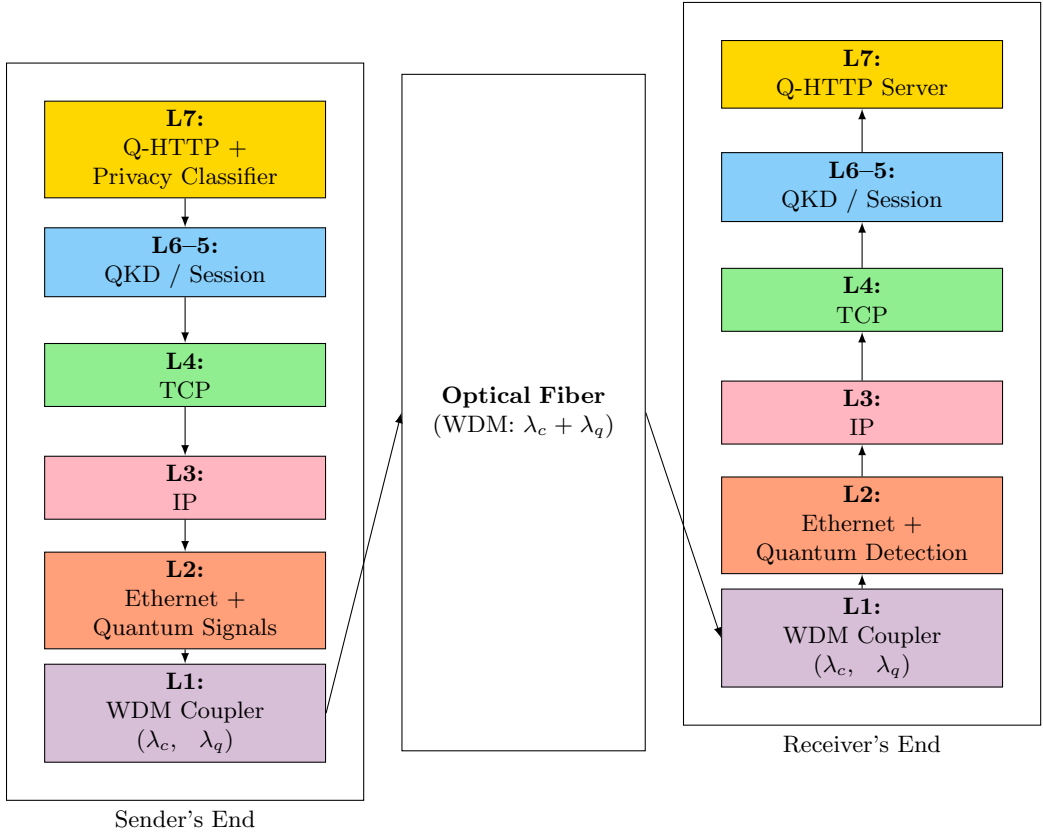
\begin{figure}[ht!]
\centering
\begin{tikzpicture}[%
  node distance=1.5cm, % Adjust spacing between layers
  >=latex, 
  every node/.style={font=\small, align=center},
  xscale=1.0,
  yscale=1.0
]

% Define colors for layers
\definecolor{layer7}{HTML}{FFD700} % Gold for Layer 7
\definecolor{layer6_5}{HTML}{87CEFA} % Light blue for Layer 6-5
\definecolor{layer4}{HTML}{90EE90} % Light green for Layer 4
\definecolor{layer3}{HTML}{FFB6C1} % Light pink for Layer 3
\definecolor{layer2}{HTML}{FFA07A} % Light salmon for Layer 2
\definecolor{layer1}{HTML}{D8BFD8} % Thistle for Layer 1

%-------------------------
% Alice's Layers (Left)
%-------------------------
\node(A7) [rectangle, draw=black, fill=layer7, text width=3.5cm] 
  {\textbf{L7:}\\ Q-HTTP +\\ Privacy Classifier};

\node(A6_5) [rectangle, draw=black, fill=layer6_5, below of=A7, text width=3.5cm] 
  {\textbf{L6--5:}\\ QKD / Session};

\node(A4) [rectangle, draw=black, fill=layer4, below of=A6_5, text width=3.5cm] 
  {\textbf{L4:}\\ TCP};

\node(A3) [rectangle, draw=black, fill=layer3, below of=A4, text width=3.5cm] 
  {\textbf{L3:}\\ IP};

\node(A2) [rectangle, draw=black, fill=layer2, below of=A3, text width=3.5cm] 
  {\textbf{L2:}\\ Ethernet +\\ Quantum Signals};

\node(A1) [rectangle, draw=black, fill=layer1, below of=A2, text width=3.5cm] 
  {\textbf{L1:}\\ WDM Coupler\\($\lambda_c,\;\lambda_q$)};

% Group Alice's layers into a box labeled "Sender"
\node [draw=black, fit={(A7) (A1)}, inner sep=0.5cm, label={[anchor=north]south:Sender's End}] {};

%-------------------------
% Single Fiber (Center)
%-------------------------
\node(F) [rectangle, draw=black, 
  minimum height=9cm,
  text width=3cm, right of=A4, xshift=3cm, yshift=-0.5cm] 
  {\textbf{Optical Fiber}\\
   (WDM: $\lambda_c + \lambda_q$)};

%-------------------------
% Bob's Layers (Right)
%-------------------------
\node(B7) [rectangle, draw=black, fill=layer7, text width=3.5cm, right of=F, xshift=3cm, yshift=4.5cm] 
  {\textbf{L7:}\\ Q-HTTP Server};

\node(B6_5) [rectangle, draw=black, fill=layer6_5, below of=B7, text width=3.5cm] 
  {\textbf{L6--5:}\\ QKD / Session};

\node(B4) [rectangle, draw=black, fill=layer4, below of=B6_5, text width=3.5cm] 
  {\textbf{L4:}\\ TCP};

\node(B3) [rectangle, draw=black, fill=layer3, below of=B4, text width=3.5cm] 
  {\textbf{L3:}\\ IP};

\node(B2) [rectangle, draw=black, fill=layer2, below of=B3, text width=3.5cm] 
  {\textbf{L2:}\\ Ethernet +\\ Quantum Detection};

\node(B1) [rectangle, draw=black, fill=layer1, below of=B2, text width=3.5cm] 
  {\textbf{L1:}\\ WDM Coupler\\($\lambda_c,\;\lambda_q$)};

% Group Bob's layers into a box labeled "Receiver"
\node [draw=black, fit={(B7) (B1)}, inner sep=0.5cm, label={[anchor=north]south:Receiver's End}] {};

%-------------------------
% Arrows for Message Flow
%-------------------------

% Alice's downward flow
\draw[->] (A7.south) -- (A6_5.north);
\draw[->] (A6_5.south) -- (A4.north);
\draw[->] (A4.south) -- (A3.north);
\draw[->] (A3.south) -- (A2.north);
\draw[->] (A2.south) -- (A1.north);

% Across the fiber
\draw[->] (A1.east) -- (F.west); % From Alice's L1 to Fiber
\draw[->] (F.east) -- (B1.west); % From Fiber to Bob's L1

% Bob's upward flow
\draw[->] (B1.north) -- (B2.south);
\draw[->] (B2.north) -- (B3.south);
\draw[->] (B3.north) -- (B4.south);
\draw[->] (B4.north) -- (B6_5.south);
\draw[->] (B6_5.north) -- (B7.south);

\end{tikzpicture}
\vspace{0.2cm}
\caption{Single-Fiber Q-HTTP Architecture. Alice’s ML Privacy Classifier (L7) determines if quantum security (QKD) is required (\texttt{label=1}). Messages flow downward through Alice's layers (L7 to L1), pass through the optical network fiber (via WDM $\lambda_c + \lambda_q$), and flow upward through Bob's layers (L1 to L7). The ``Sender" and ``Receiver" groups encapsulate the respective layers.}
\label{fig:layers}
\end{figure}%diagram arch
The following gives a brief description of the layers presented in fig(\ref{fig:layers}), and how it's modified from current classical framework. 
\begin{enumerate}
    \item \textbf{L7: Q-HTTP + Privacy Classifier}:
     In the classical OSI model, the application layer (Layer 7) typically employs HTTP/HTTPS. Q-HTTP (as described in fig(\ref{fig:http-quant})) extends HTTP to request quantum (QKD-based or otherwise) encryption, while a privacy classifier assigns a label indicating whether an email requires quantum encryption.

     \item \textbf{ L6–5: QKD / Session}:
     In classical OSI, the session layer (Layer 5) handles sessions, and the presentation layer (Layer 6) manages data formats, compression, and encryption. Modern stacks often merge both into the application layer. Here, they are combined into a single ``Session + QKD" layer that establishes quantum-derived keys (or more complex quantum secure direct communication protocols)  and maintain session state.

     \item \textbf{L4: TCP}:
     This layer should remain rather similar to classical framework.  TCP handles reliable data delivery, packet reordering, etc.

     \item \textbf{L3: IP}:
     This layer is also almost unchanged as this is the routing of the packets based on IP. Now, we would have packets with quantum encrypted states or quantum information, but this doesn't change the way routing would changed. Only change would be to direct packets having quantum encryption to quantum gateways. 

     \item \textbf{L2: Ethernet + Quantum Signals}:
     Ethernet is the standard data-link protocol for framing and addressing. In this hybrid model, it handles both classical traffic and quantum signals, with specialized photonic channels or hardware managing the latter.

     \item \textbf{L1: WDM Coupler $(\lambda_c, \lambda_q)$}:
     This layers mainly manages the transmission medium and signaling methods.  We propose WDM can be used to manage both quantum and classical information just at different wavelengths using same optical fiber.

\end{enumerate}

\section{Methodology}
\label{Sec:methods}
In this section, we will go over the details of the experimental setup. We will go in detail about the NLP methods mentioned in fig(\ref{fig:nlp-struct}) concerning the discussion of quantum augmented networks. We will present the methodology used for the preparation of the dataset, and high-level diagrams of the models used (Logistic Regression, CNN, LSTM, and BiLSTM). 

\subsection{Dataset Description}
\label{Sec:Dataset}
One of the major challenges for this study was curating an appropriate dataset. For this purpose, we start with the Enron Email dataset \cite{enron_email_dataset}. This dataset consists of $587,000$ uncategorized emails. We do some private information injections into some of the emails to create a new dataset which consists of $20,000$ emails out of which $10,000$ are general non-private emails and $10,000$ are emails injected with private information. The injection is done through the use of several layers of regular expressions in the following way, 
\begin{enumerate}

    \item Randomly sample $20,000$ emails from the original Enron email dataset.
    \item Use of regular expressions and a list of contextual words (such as ``money'' can be a good identifier for financial emails) to identify the context of the emails, i.e., if any private information injection can be done. 
    \item In this study, we only consider communication scenario between two users. The definition of privacy is very broad, however for this study we limit it to \textit{financial}, and \textit{personal} information. Therefore, we inject the private information from one of the categories,
    \begin{itemize}
        \item SSN
        \item Email address
        \item Password
        \item Credit Card
        \item Account balance
        \item Transaction ID
    \end{itemize}
    This list can be further extended to more real word scenarios, and thus make the dataset more personalized for different use cases, such as healthcare, employee-employer relationships, governmental communications, etc. 
    \item We label all of the emails with any of the above injections as $1$. Rest of the emails are kept as label $0$.
\end{enumerate}
\begin{tcolorbox}[colback=gray!10, colframe=black, title=Email Example]
\begin{verbatim}
Sender: mary23@hotmail.com
To: nitin@gmail.com
Sub: Sensitive information asked earlier. 

Dear Nitin,
 I am attaching my SSN in this email as discussed in our previous conversation,
Let me know if you need anything else, 
123-456-7890.

Thanks.
Best,
Mary
\end{verbatim}
\end{tcolorbox}

\begin{figure}[h!]
    \centering
    \includegraphics[width=0.8\linewidth]{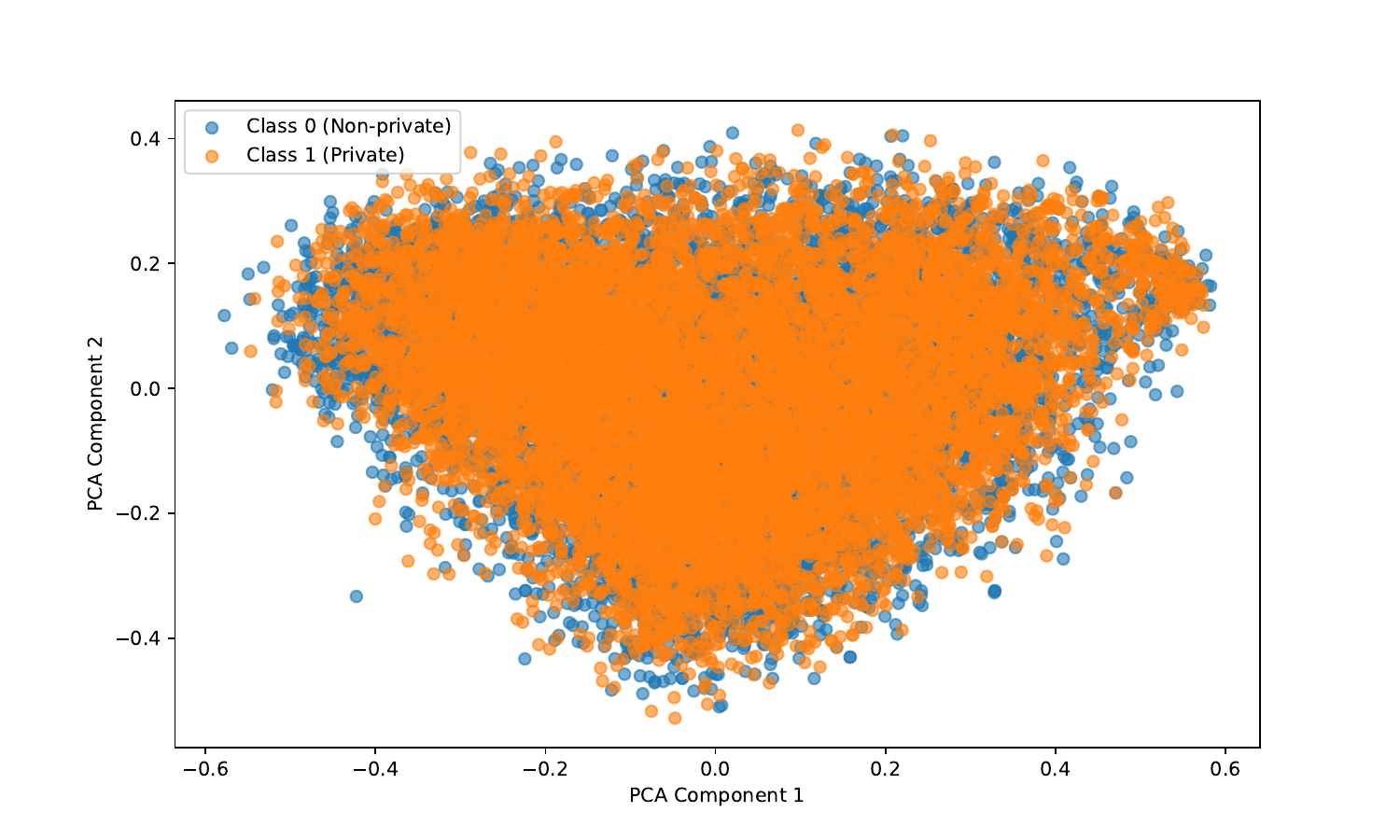}
    \caption{PCA Visualization of Email Embeddings.}
    \label{fig:pca-visualization}
\end{figure}

\begin{figure}[h!]
    \centering
    \includegraphics[width=0.8\linewidth]{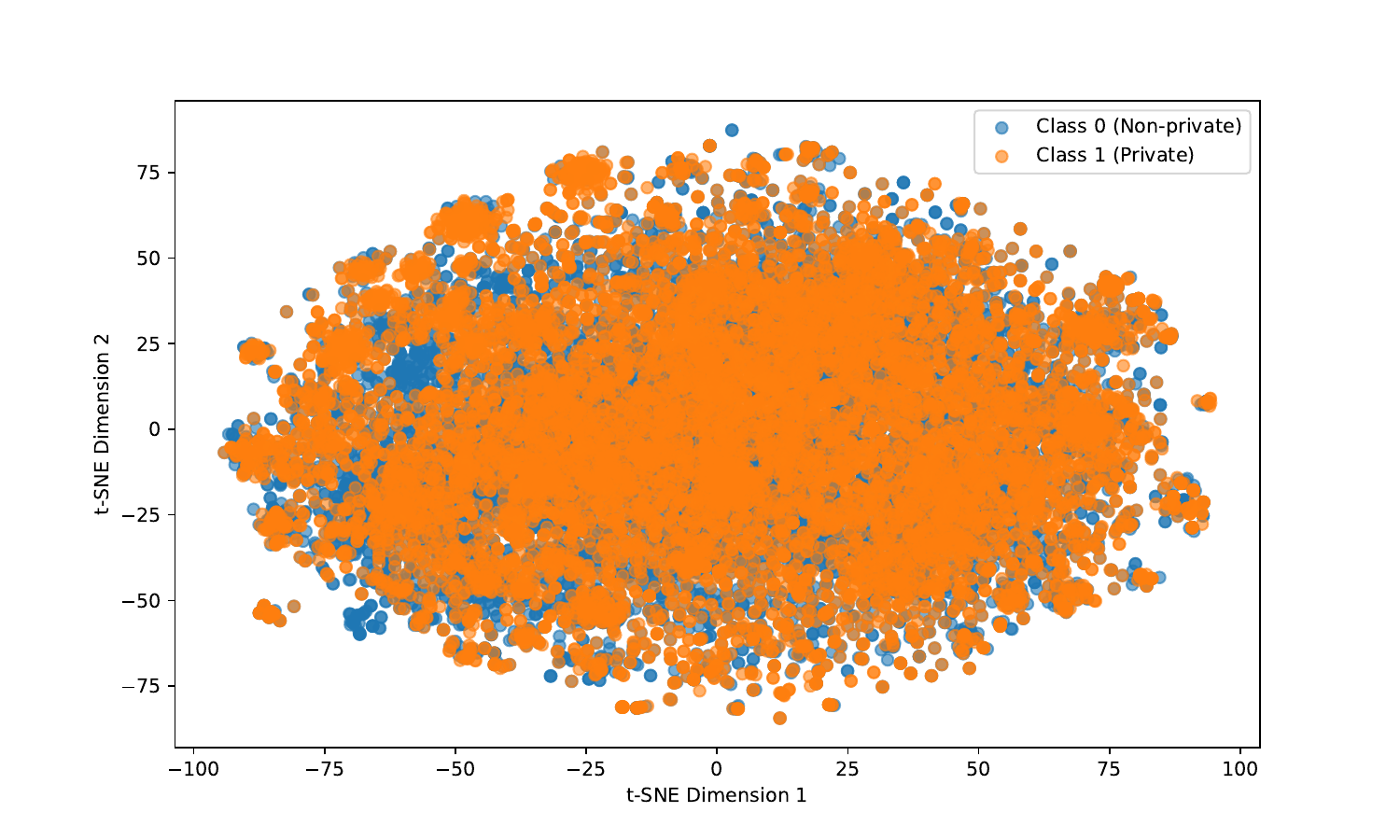}
    \caption{t-SNE Visualization of Email Embeddings. }
    \label{fig:tsne-visualization}
\end{figure}

Out of this dataset, $80\%$ was used to train and validate the model and $20\%$ for testing and evaluation. We present clustering of the dataset to visualize the similarity between the two classes, private and non-private emails. We use PCA analysis and t-SNE reduction to visualize the similarity between the two classes in our dataset. From fig(\ref{fig:pca-visualization}) and fig(\ref{fig:tsne-visualization}), we can see that both of the classes are similar and have overlapping content, this is important to check as we do not want any biases while training any of the models.

\subsection{Model Architecture}

In this subsection, we detail the implementation of the various machine-learning models utilized for detecting privacy labels in emails. These models were trained on the dataset described in Sec[\ref{Sec:Dataset}]. Each model leverages different learning paradigms and architectures to predict whether a given email should be classified as private or non-private. The models implemented include Logistic Regression, Convolutional Neural Networks (CNNs), Long Short-Term Memory (LSTM) networks, and Bidirectional LSTM (BiLSTM) networks. Each of these models contributes to the overall analysis of privacy classification in emails, allowing us to evaluate their respective performance and determine the optimal approach for quantum-classical augmented networks. Tab(\ref{tab:models}) provides a summary of all the models used in this study.

\begin{table}[ht!]
\centering
\renewcommand{\arraystretch}{1.5} % Increase row height
\begin{tabular}{|p{2.5cm}|p{4cm}|p{3.5cm}|p{4cm}|}
\hline
\textbf{Model} & \textbf{Architecture Details} & \textbf{Key Parameters} & \textbf{Training Details} \\ \hline

\textbf{Logistic   Regression}   & 
TF-IDF Vectorizer (5000 features) + Logistic Regression & 
\texttt{max\_features=5000} (TF-IDF) & 
Not Applicable \\ \hline

\textbf{CNN} & 
Embedding Layer $\rightarrow$ Convolutions $\rightarrow$ MaxPooling $\rightarrow$ Dropout $\rightarrow$ FC $\rightarrow$ Sigmoid & 
\texttt{vocab\_size=5000}, \texttt{num\_filters=100}, Dropout: $0.4$ & 
\texttt{lr=0.001}, Optimizer: Adam, Loss: BCE \\ \hline

\textbf{LSTM} & 
Embedding Layer $\rightarrow$ LSTM $\rightarrow$ BatchNorm $\rightarrow$ Dropout $\rightarrow$ FC $\rightarrow$ Sigmoid & 
\texttt{vocab\_size=5000}, \texttt{lstm\_units=128}, Dropout: $0.5$ & 
\texttt{lr=0.0005}, Batch Size: $64$ \\ \hline

\textbf{BiLSTM} & 
Embedding Layer $\rightarrow$ BiLSTM $\rightarrow$ Dropout $\rightarrow$ FC $\rightarrow$ Sigmoid & 
\texttt{vocab\_size=5000}, \texttt{lstm\_units=64}, Dropout: $0.5$ & 
\texttt{lr=0.0001}, Batch Size: $64$ \\ \hline

\end{tabular}
\vspace{0.2cm}
\caption{Summary of models and their parameters used for privacy classification.}
\label{tab:models}
\end{table}

We need to use more complex models such as LSTM, BiLSTM to identify long-range dependencies which might not be identified using unigrams, bigrams, or trigrams. An example of such would be as follows.

Here we can clearly see that the sender mentioned in the earlier part of the email that she's attaching SSN in this email, and the actual SSN is at the end of this email. This might be missed if we only use n-grams, and this is reflected in the results for the baseline model. 

\subsubsection*{Logistic Regression}
Logistic Regression was used as a baseline model. We extracted textual features from emails using TF-IDF (Term Frequency-Inverse Document Frequency) vectorization with a maximum of 5000 features. The processed feature vectors were then passed to a standard Logistic Regression classifier. This approach, while simplistic, provides a robust baseline for comparison with more complex neural models. The following is a 

\begin{itemize}
    \item \textbf{Feature Extraction}: TF-IDF Vectorizer (\texttt{max\_features=5000}) using unigram, bigrams, and trigrams.
    \item \textbf{Classifier}: Logistic Regression.
\end{itemize}

\subsubsection*{Convolutional Neural Network (CNN)}
CNNs are designed to capture spatial and sequential relationships in data. Here, we utilized a one-dimensional CNN to extract features from embedded word vectors. The model architecture comprises an embedding layer followed by multiple convolutional layers with different filter sizes, max-pooling, a fully connected layer, and a sigmoid activation for binary classification.
\subsubsection*{Long Short-Term Memory (LSTM)}
LSTM networks are recurrent models that effectively capture long-term dependencies in sequential data. The LSTM model comprises an embedding layer, an LSTM layer with batch normalization, dropout, a fully connected layer, and a sigmoid activation for binary classification.

\subsubsection*{Bidirectional LSTM (BiLSTM)}
The BiLSTM model extends the LSTM architecture by incorporating both forward and backward LSTM layers. This enables the model to capture contextual information from both directions. The architecture includes an embedding layer, bidirectional LSTM, dropout, a fully connected layer, and a sigmoid activation.

\begin{figure}[h!]
    \centering
    \includegraphics[width=0.7\linewidth]{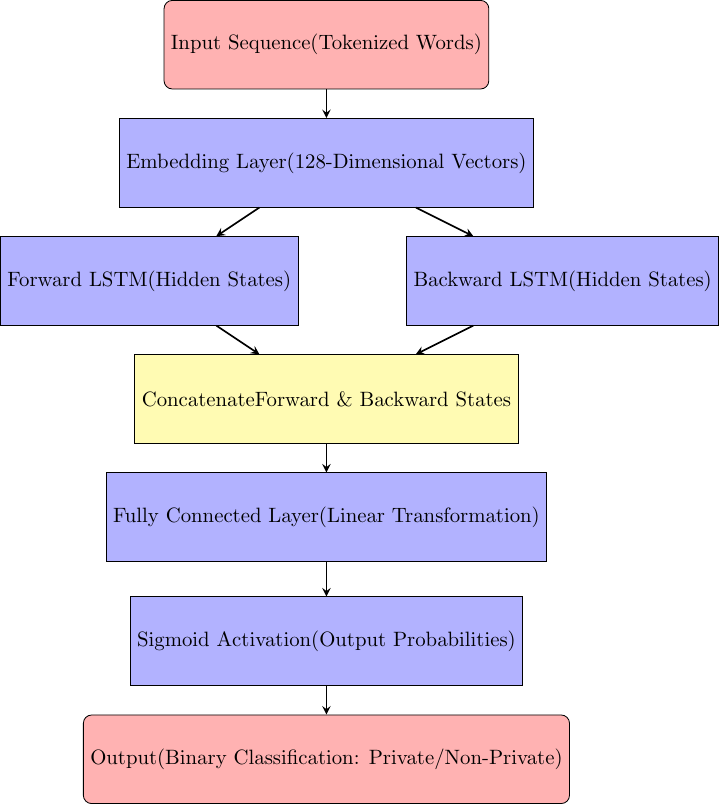}
    \vspace{0.3cm}
    \caption{Architecture of the BiLSTM model used in this experiment.}    
    \label{fig:bilstm}
\end{figure}

\section{Results}
\label{Sec:result}
In this section, we will present the results of the several models mentioned in Sec[\ref{Sec:methods}]. We will present the classification reports for all the models and the confusion matrices associated for each. 
\subsection*{Logistic Regression}
Logisitic regression is a basic model and serves as a baseline for comparison of the performance of the more complex models. For $5000$ features, we get an accuracy of $88.10\%$. This is a fairly good performance for baseline model. 

\begin{table}[h!]
    \centering
    \caption{Classification Report for LR Model}
    \begin{tabular}{@{}lccccr@{}}
        \toprule
        Class & Precision & Recall & F1-Score & Support \\ \midrule
        0     & 0.86      & 0.95   & 0.91     & 2013    \\
        1     & 0.95      & 0.84   & 0.91     & 1987    \\ \midrule
        Accuracy & \multicolumn{3}{c}{0.90} & 4000    \\ \midrule
        Macro Avg & 0.91   & 0.90   & 0.90     & 4000    \\
        Weighted Avg & 0.90 & 0.90   & 0.90     & 4000    \\ \bottomrule
    \end{tabular}
    \label{fig:lr_class}
\end{table}

We see that the recall of the private class (label $1$) is just $84\%$ for the private class in tab(\ref{fig:lr_class}). This is good, but we need to ensure, we have high recall and precision scores for private class. The overall accuracy of this basline model, (which uses n-grams for vectorization) was $90\%$ which is a good baseline, however the low recall values for private class can be a big issue, and shows that this model might not be capturing several long-range dependencies for private class emails, and thus we need to analyze the performance of more complex models. 

\begin{figure}[h!]
    \centering
    \includegraphics[width=0.6\linewidth]{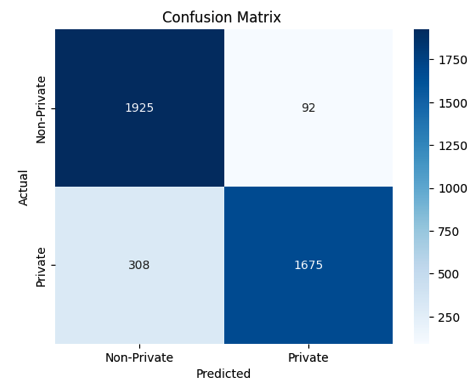} % 
    \caption{Confusion matrix obtained for Logistic Regression model as described in tab{\ref{tab:models}}.}
    \label{fig:lr_cm}
\end{figure}

 The confusion matrix (fig(\ref{fig:lr_cm}) shows that the false negative rates are quite high. A large section of private emails are classified as non-private. This is not good for any future practical applications. 
\subsection*{CNN Model}
CNN model serves as the basis of performance comparison between the linear logistic regression models and neural models. There is a significant increase in the accuracy to $94.63\%$. However, to analyze the overall performance of the CNN model, we have to look at the classification report.

\begin{table}[h!]
    \centering
    \caption{Classification Report for CNN Model}
    \begin{tabular}{@{}lccccr@{}}
        \toprule
        Class & Precision & Recall & F1-Score & Support \\ \midrule
        0.0   & 0.93      & 0.97   & 0.95     & 1987    \\
        1.0   & 0.97      & 0.93   & 0.95     & 2013    \\ \midrule
        Accuracy & \multicolumn{3}{c}{0.95} & 4000    \\ \midrule
        Macro Avg & 0.95   & 0.95   & 0.95     & 4000    \\
        Weighted Avg & 0.95 & 0.95   & 0.95     & 4000    \\ \bottomrule
    \end{tabular}
\end{table}

We see a significant increase in both the precision and the recall of the model compared to the earlier linear model. We see a great improvement in recall of private class from $84\%$ to $93\%$ and we also see that the precision for predicting private emails is $97\%$, which is great for any future applications in augmented network scenarios, as mentioned in fig(\ref{fig:nlp-struct}).

\begin{figure}[h!]
    \centering
    \includegraphics[width=0.6\linewidth]{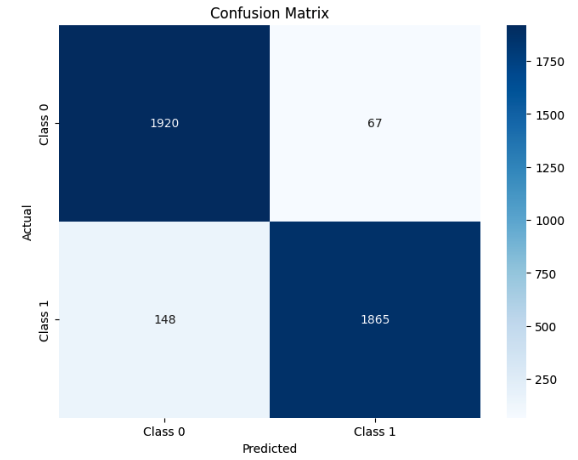} % 
    \caption{Confusion matrix obtained for CNN model as described in tab{\ref{tab:models}}.}
    \label{fig:cnn_cm}
\end{figure}

We also see that the false negative for the private class is almost half that of the linear model. This is very important for practical purposes as classifying a non-private message as private and offering a higher level of security through quantum encryption is better than risking a classical encryption of a highly private message over any future augmented networks. 

\subsection*{LSTM Model}
We expect the LSTM model to perform better than the CNN model due to its ability to capture long-range dependencies, contextual recognition, etc. The accuracy of the model was found to be $95.4\%$, which is a significant improvement over CNN model. 

\begin{table}[ht]
    \centering
    \caption{Classification Report for LSTM Model}
    \begin{tabular}{@{}lccccr@{}}
        \toprule
        Class      & Precision & Recall & F1-Score & Support \\ \midrule
        Class 0    & 0.99      & 0.91   & 0.95     & 2016    \\
        Class 1    & 0.92      & 0.99   & 0.96     & 1984    \\ \midrule
        Accuracy   & \multicolumn{3}{c}{0.95} & 4000    \\ \midrule
        Macro Avg  & 0.96      & 0.95   & 0.95     & 4000    \\
        Weighted Avg & 0.96    & 0.95   & 0.95     & 4000    \\ \bottomrule
    \end{tabular}
\end{table}

Based on the full classification report, we see that the performance of the LSTM model is better than the CNN model. We notice that the model predicts non-private emails with $99.75\%$ accuracy. Moreover, there's no false negatives for private class, which is fairly a good result for practical purposes. 
\begin{figure}[h!]
    \centering
    \includegraphics[width=0.6\linewidth]{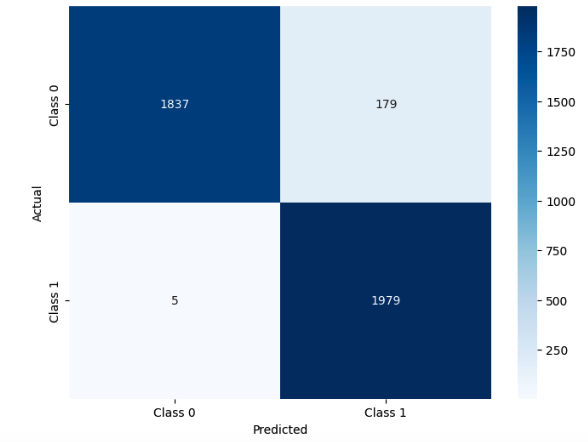}
    \caption{Confusion matrix obtained for LSTM model as described in tab{\ref{tab:models}}.}
    \label{fig:lstm_cm}
\end{figure}

\subsection*{BiLSTM Model}
BiLSTM model, in theory, should be able to capture more context than a LSTM model due to the bi-directional component. This is reflected in the increased accuracy of $94.17\%$. This is similar to the CNN model, but is a bit lower than the LSTM model. 

\begin{table}[h!]
    \centering
    \caption{Classification Report for BiLSTM Model}
    \begin{tabular}{@{}lccccr@{}}
        \toprule
        Class      & Precision & Recall & F1-Score & Support \\ \midrule
        0.0        & 0.93      & 0.96   & 0.94     & 1994    \\
        1.0        & 0.95      & 0.93   & 0.94     & 2006    \\ \midrule
        Accuracy   & \multicolumn{3}{c}{0.94} & 4000    \\ \midrule
        Macro Avg  & 0.94      & 0.94   & 0.94     & 4000    \\
        Weighted Avg & 0.94    & 0.94   & 0.94     & 4000    \\ \bottomrule
    \end{tabular}
\end{table}

From all of the different classification reports, we notice that the LSTM model outperforms CNN and the BiLSTM model. 

\begin{figure}[h!]
    \centering
    \includegraphics[width=0.6\linewidth]{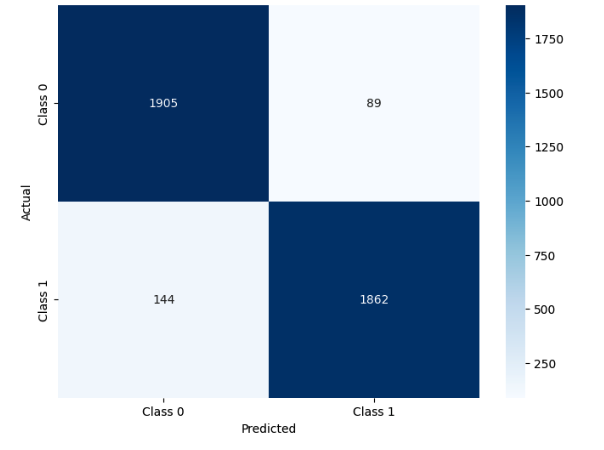} % Replace with your image file
    \caption{Confusion matrix obtained for BiLSTM model as described in tab{\ref{tab:models}}.}
    \label{fig:bilstm_cm}
\end{figure}
From the confusion matrix, we notice that the false negative rate for the private class is comparable to the CNN model.The result is quite acceptable for all the models which shows average f-1 scores of $94\%$, thus showing that the classification task can be carried out in a correct manner using any of these three models. All these models are shown to be quite robust with the parameters mentioned in tab(\ref{tab:models}). Further experiments can be conducted to explore for more complex classifications, i.e., context-based classification.

\begin{figure}[h!]
    \centering
    \includegraphics[width=\textwidth]{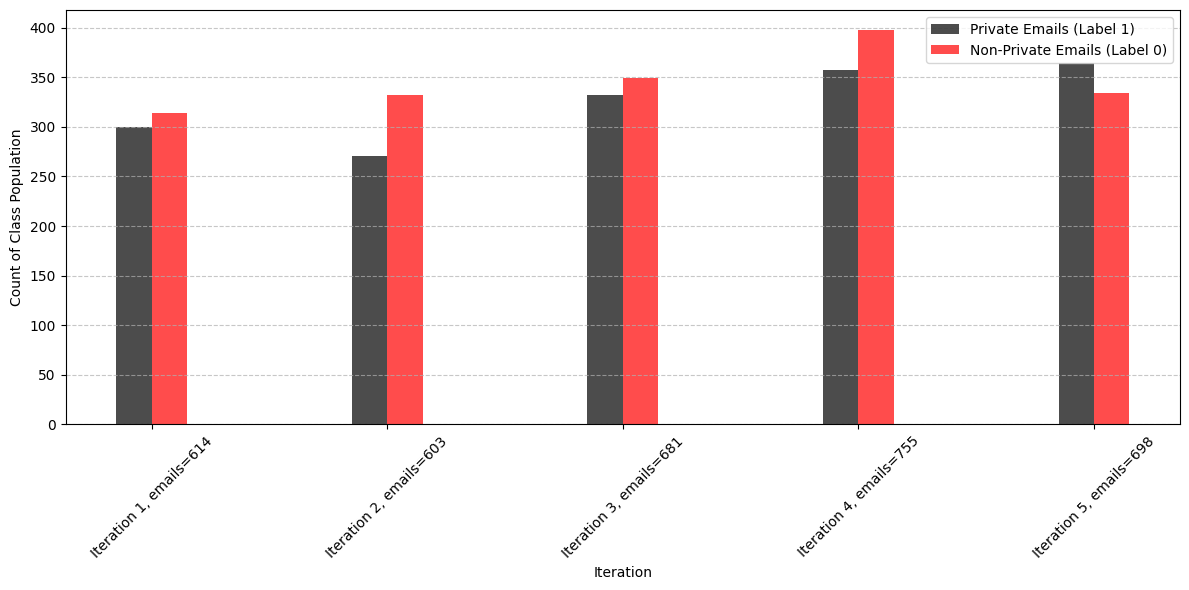}
    \caption{Comparison of private vs non-private email communications to represent a communication scenario of user sending emails over a communication network. Red bars represents the number of non-private email, and black bars represents the number of private emails in given network iteration.}
    \label{fig:quant-ratio}
\end{figure}

The main idea is to reduce the load on quantum networks by limiting the amount of quantum encryption to only the messages having private contents. To assess this, we randomly sample $500$-$800$ emails from our dataset to represent email communications over a network and keep the count of privacy labels for these emails. We do this sampling fives times, and present the ratio of messages private emails to non-private ones in fig(\ref{fig:quant-ratio}).

%this is a placeholder, will replace with actual diagram soon

From the above figure, we can see that for each of the five iterations, the split between private and non-private emails are almost $50\%-55\%$. In a purely quantum network scenario, we would need to transmit all of these emails (as listed for each iteration on the x-axis) using quantum encryption. However, when using our proposed method of assigning a privacy label and selectively using quantum encryption, we can see that we would only need to use about $50\%$ or less quantum resources from a purely quantum network while still achieving an acceptable level of security through the use of quantum encryption for private content, and classical encryption techniques for non-private communications. Thus, the security offered is higher than that of a classical network and similar to a purely quantum network for this quantum-augmented network setup. At the same time, it is much more resource-efficient than a fully quantum network as we are using quantum encryption selectively through the use of the proposed privacy classifier. This can further be extended to approximate the reduction in the amount of photons or other relevant physical resources needed for our quantum-augmented network over a purely quantum network. It is intuitive from the above figure that there is a significant reduction in using quantum encryption for our example case. This study serves as a baseline to show that quantum-augmented networks serve as a more resource-efficient solution to purely quantum setups while fulfilling a higher security output than that of a classical network scenario.

\section{Conclusions}
\label{Sec:conclusion}
The main focus of this work was to develop a machine learning framework for privacy detection in any textual communication over a quantum-augmented network. The idea is to limit the use of quantum resources by only encrypting communications with highly private content. This would significantly improve the efficiency of quantum communication, as quantum resources have significantly higher costs than any classical network resource. In this work, we curated a custom dataset of 20,000 emails, with two classes: private and non-private, each with 10,000 emails. We then trained the logistic regression model, CNN model, LSTM, and BiLSTM models to compare their performance for this classification task. Our results showed that the LSTM models outperform the CNN, the BiLSTM models, and the baseline linear regression model. To understand the significance of this proposed approach, we took a mock network scenario by sampling a random number of emails to show that selectively encrypting messages will be more efficient approach in quantum-augmented network scenarios.  Future work will focus on modifying the BiLSTM model to include attention weights to identify specific private content in the messages which are labeled private, which will help users get more insights about why a specific label is attached to a given message.   More developments would require privacy classifications with images and other formats to cover private information being sent across a network. This is one of the initial steps towards a fairly new approach of quantum-augmented networks instead of a stand-alone quantum network. 

\section*{Acknowledgment}
This work is partly sponsored by the National Science Foundation (NSF) awards numbers 2324924 and 2324925.

% References
\bibliography{bibliography} % bibliography data in report.bib
\bibliographystyle{spiebib} % makes bibtex use spiebib.bst

\end{document}